\documentclass[12pt]{article}
\usepackage{cyr}
\usepackage{epsf}
\usepackage{pazh}

\usepackage{amsmath}
\usepackage{amssymb}
\usepackage{graphicx}
\usepackage{indentfirst}
\usepackage{lscape}
\usepackage{afterpage}

\tightenlines

\def\*{$^{*}$}
\def\Á{$^{\mbox{\small Á}}$}
\def\Â{$^{\mbox{\small Â}}$}
\def\×{$^{\mbox{\small ×}}$}
\def\Ç{$^{\mbox{\small Ç}}$}
\def\Ä{$^{\mbox{\small Ä}}$}
\def\ÅÒÇÓ{ÜÒÇ~Ó$^{-1}$}
\def\ÅÒÇÓÍ{ÜÒÇ~ÓÍ$^{-2}$~Ó$^{-1}$}

\sloppypar

\begin{document}
\baselineskip 21pt

\title{On the mass and the density of stellar disk of M33.}

\author{\bf \hspace{-1.3cm}\copyright\, 2011  \ \
A. S. Saburova \affilmark{1*}, A. V. Zasov\affilmark{1**}}

\affil{ {\it Sternberg Astronomical Institute of Moscow State University}$^1$}

\vspace{2mm}

\sloppypar \vspace{2mm} \noindent The disk surface density of the nearby spiral galaxy M33
is estimated assuming that it is marginally stable against gravitational perturbations.
For this purpose we used the radial profile of line-of-sight velocity dispersion of the
disk planetary nebulae obtained by Ciardullo et al. (2004). The surface density profile we
obtained is characterized by the radial scalelength which is close to the photometrical
one and is in a good agreement with the rotation curve of M33 and with the mass-to-light
ratio corresponding to the observed color indices. However at the galactocentric distance
$r>7$ kpc the dynamical overheating of the disk remains quite possible. A thickness of the
stellar disk of M33 should increase outwards. The dark halo mass exceeds the  mass of the
disk at $r>$ 7 kpc. The obtained radial profile of the disk surface density and the radial
gradient of $O/H$ are used to calculate the effective oxygen yield $Y_{eff}$  in the frame
of the instantaneous recycling approximation. It is shown that $Y_{eff}$ increases with
radius which may indicate that the role of accretion of metal-poor gas in the chemical
evolution of interstellar medium decreases outwards.
\\
\noindent {\bf Key
words:\/} galaxies, galactic disks, gravitational stability of stellar disks\\
\vfill \noindent\rule{8cm}{1pt}\\ {$^{*}$ e-mail: $<$saburovaann@gmail.com$>$}\\
{$^{**}$ e-mail: $<$a.v.zasov@gmail.com$>$}\\

\clearpage

\section{ Introduction }
M33 is the late-type spiral galaxy of rather small size which belongs to the Local Group.
Optical radius of the galaxy inside of $25^m/\Box''$ isophote in B-band is close to 10
kpc. Due to the large angular size and the proximity to the Galaxy M33 is a perfect object
for a detailed study of the density distribution of luminous and non-luminous matter. Of
particular interest to this galaxy is the dark halo which includes a significant fraction
of total mass within the optical borders. The indirect manifestation of the massive halo
is the continued growth of the rotation curve at radial distances $r>2h$, where h is a
photometric disk radial scalelength (Persic et al., 1996). However the shape of the slowly
rising rotation curve allows different ways to decompose the curve into the disk and dark
halo components, which creates a problem of their mass ratio estimation. The attempts to
decompose the galaxy into disk and halo components were done by Corbelli, Salucci (2000)
and Corbelli (2003), which confirmed the domination of the dark halo within the optical
radius. Assuming the disk radial scalelength to be equal to the K-band photometric
scalelength, Corbelli (2003) came to conclusion that in the best fit model the rotation
curve may be explained by the dark halo, which mass is about ten times  higher than  the
mass of stellar disk, or at least five times higher than the disk baryonic mass
(stars+gas)  within two optical radii $r_{25}$. Independent disk mass estimation based on
the condition for the existence of the wave spiral structure in M33 led to the dark halo
mass fraction $M_h/M_d=0.83$ inside the optical borders (see Athanassoula et al. 1987).

Later Ciardullo et al. (2004) estimated the disk mass of M33 by analyzing the measured
velocity dispersion of planetary nebulae belonging to the stellar disk. By introducing
some simplifying assumptions, these authors evaluated the vertical component of the
velocity dispersion $\sigma_{z}$ for the galactocentric distances $r=$1.2--8.5 kpc to
obtain the surface density of the equilibrium disk.  The disk half thickness $z_*$ (the
scaleheight of planetary nebulae) was considered to be constant with radius (quite
arbitrary it was assumed that $z_*$ = 175 pc). The gravitational stability condition of a
thin disk was superposed as an additional requirement. The model of the galaxy obtained by
Ciardullo et al. (2004) differs much from that of Corbelli (2003): the disk radial
scalelength found by Ciardullo et al. (2004) is twice as high as the optical K-band scale, which
leads to conclusion that the mass-to-light ratio grows with galactocentric radius: $M/L_V$
should be about five times higher in the periphery than inside of $r\approx$ 2 kpc. The
physical reason for this increase remains enigmatic.

 The alternative way to measure a disk mass can be based on the assumption
 that the disk self-gravity leads the disk  to be close to the marginal
 stability state against the gravitational perturbations (this approach was first proposed
 by Zasov, 1985 and Morozov, Zasov, 1985 and later developed by Bottema, 1993). In general case, when marginal stability
 condition does not hold, the resulting density and mass estimates can be treated as
 the upper limits. There are some arguments supporting the idea that the disks of
 spiral galaxies
 are usually close to marginal stability. These are the absence of systematic deviation between the
  $M/L$ values found for spiral galaxies under this assumption and from the photometrical models,
  and also the existence of correlation between the relative thickness of edge-on disks and
  the relative mass
  of their dark halo (see the discussion in Zasov et al. 2002, Zasov et al. 2011). The most promising way
to construct models of marginally stable disks is to use the numerical 3D dynamical
models. The numerical modeling of M33 disk confirms that the disk mass is low in
comparison with the dark halo mass: if to admit that the photometrical
 radial scalelength is close to the disk density scalength, then the velocity dispersion of
 planetary nebulae agrees with the model where the disk mass does not exceed
 one-third of the total mass inside of four radial scalelengths (Fridman, Khoperskov, 2011).

In the current paper we re-estimate the upper limit of the disk surface density using the
line-of-sight velocity dispersion of planetary nebulae at different galactocentric
distances $r$  obtained by Ciardullo et al. (2004). As it follows from the analysis of
stellar composition of M33, most stars of the galaxy have a large age exceeding 2-3
billion years (see e.g. Williams et al., 2009). As far as the planetary nebulae are
associated with old stars (red giants), their velocity dispersion can be ascribed to stars
which constitute the main body of the disk. Notice that the the mean velocity dispersion
of the disk red giants is  $\sim$ 24 km/s (Hood et al. 2009), -- in a good agreement with
that of planetary nebulae.

Thus, the initial assumption is that the disk of M33 (at least inside of the radius
covered by the velocity dispersion measurements) has the surface density which is close to
the threshold for the gravitation stability (i.e. the disk is assumed to be in the
equilibrium and marginally stable state). Below we use the modified analytical stability
criterion and  consider the disk radial scalelength  and mass as a priori unknown
parameters. The local disk density values found in this work were checked for
compatibility with the rotation curve of the galaxy and with the photometrically
determined radial profile of $M/L$ for stellar disk as well as with the available
estimates of radial distribution of oxygen $O/H$. Following Ciardullo et al. (2004), we
take the distance to the galaxy D = 0.94 Mpc

\section{The stability criterion and the role of gas component}
The constraints on the surface density of marginally stable disk can be found when the
radial velocity dispersion of stars or planetary nebulae in a disk  (that is the velocity
dispersion along the radial coordinate $r$) $c_r$ is known:
\begin{equation}
\label{formula1} \sigma = \frac {c_r \kappa} {3.36GQ_c},
\end{equation}
where $Q_c=c_r/c_T$ is the Toomre parameter of stability, $c_{T}$ is the Toomre critical
velocity dispersion, $\kappa$ is epicyclical frequency defined as a function of the
angular velocity $\Omega $ and its derivative:
\begin{equation}
\label{formula3} \kappa =2 \Omega\cdot\sqrt {1+(r/2\Omega)(d\Omega/dr)}.
         \end{equation}
The angular velocity at a given galactocentric distance  was taken from the rotation curve
of M33  obtained by Corbelli (2003). As the input data for stellar velocity dispersion we
have used the radial profile of the observed line-of-sight velocity dispersion of the disk
planetary nebulae given in Ciardullo et al., 2004 (see their Fig. 8).

Radial velocity dispersion is related to the line-of-sight dispersion $c_{obs}$ through an
obvious equation:
\begin{equation}
\label{formula4} c_{obs}(r) = (c_z ^2\cdot cos^2 (i)+ c_{\phi}^2\cdot sin^2  (i)\cdot
cos^2(\alpha)+ c_{r}^2\cdot sin^2(i)\cdot sin^2(\alpha))^{1/2},
\end{equation}
where $c_z$, $c_{\phi}$, $c_{r}$ are the vertical, azimuthal and radial components, $i$ is
the disk inclination, $\alpha$ is the angle in the disk plane between the radius-vector of
a planetary nebula and the major axis. To separate the components of the velocity
dispersion one can introduce two additional conditions: $ c_{r} = 2\Omega \cdot c_{\phi}
/\kappa $ (Lindblad formula for the epicyclical approximation) and $ c_z=m\cdot c_{r} $,
where $m$ is a coefficient assumed to be constant with radius. Both the available
measurements of velocity dispersion of the galactic disks (Shapiro et al., 2003) and the
results of numerical modeling (see e.g. Zasov et al., 2008) show that in most cases  $m$
lays within the range 0.4-0.7.

If the variation of the line-of-sight velocity dispersion along the disk major axis is
known, then, taking $\alpha$ =0 in (3), we have:
\begin{equation}
\label{formula4à} c_r = c_{obs}(m^2cos^2(i) +(\kappa/2\Omega)^2 sin^2(i))^{-1/2}
\end{equation}
For the assumed inclination of the disk  (i=$56^0$) the value of $c_r$ weakly depends on
$c_z$: the transition from $m$= 0.4 to $m$ = 0.7 changes  $c_r$ by no more than 15\%, so
we restrict ourselves to the case of $m$= 0.4 below. As it is followed from H\small{I}
observations, the gaseous disk of M33 is warped in the outer part ($r>30$'), and, as the
result, its inclination and the position angle vary with the galactocentric distance
(Corbelli,  Schneider 1997).  However this variation is significant only for the last
point of the velocity dispersion profile of planetary nebulae. Therefore we assume the
inclination and position angle of the disk to be fixed.

For the axisymmetric perturbations of marginally stable thin disk the Toomre parameter $
Q_c =1$. Non-axisymmetric perturbations require higher velocity dispersion to stabilize
the disk while the finite thickness of the disk, by contrast, makes it more stable. In
general case, the critical value of $ Q_c$ at a given radius depends in a complicated way
on the geometrical and kinematical parameters of a disk as well as on the initial
conditions of its dynamical evolution. A large number of 3D equilibrium models of
marginally stable disks  demonstrate that in the absence of massive bulge the critical
value of Toomre parameter increases with the radius, and this variation may be
approximated by parabola:
\begin{equation}
\label{formula5} Q_c(r/h) = A_0+A_1\cdot  (r/h)+A_2\cdot  (r/h)^2
  \end{equation}
where $A_0=1.25$,  $A_1=-0.19$,  $A_2=0.134$ and $r/h$  is the radial distance expressed
in the disk scalelength units (Khoperskov et al., 2003).

M33 is a gas-rich galaxy, where the ratio of gas to stellar surface densities exceeds 25\%
beyond  $r\approx4$ kpc, and the ratio of total masses of gaseous and stellar disks inside
of the optical borders exceeds unity (see Corbelli (2003)). Hence the gas may play a
significant role being a cold collisional component of the disk. The influence of a gas on
a disk stability was investigated by different authors under certain simplifications. Here
we use the results of analytical computation of stability of two-component models of a
thin disk following Rafikov (2001) and Morozov, Khoperskov (2005).

Rafikov (2001) considered the Toomre parameters for purely gaseous and stellar components:
$Q_{*}=\frac{\kappa c_{*}}{\pi G \sigma_{*}}$, $Q_{gas}=\frac{\kappa c_{gas}}{\pi G
\sigma_{gas}}$ and the ratio  $\nu=c_g/c_*$ of velocity dispersions of gas and stars. In
his paper Rafikov presented the diagram $1/Q_{gas}-1/Q_{*}$ for marginally stable
stellar-gaseous disks for different values of $\nu$. In turn, Morozov and Khoperskov
derived the relationship between the radial velocity dispersion needed for stability of
stellar-gaseous disk as a whole as a function of  the ratios of velocity dispersions and
surface densities of stellar and gaseous disk components.

Using these two papers, we found the coefficients corresponding to the observed ratios of
the surface densities and velocity dispersions of stars and gas, which describe how the
critical  Toomre parameter should be changed to account for the gas in M33. These
coefficients, lying in the range  1.1 - 1.9, were applied to the values of $Q_c$ obtained
for pure stellar disk from the equation (5). Thereby it is assumed that the corrections
for the gas influence on the disk stability, obtained within the framework of simple
analytical models, are valid for the disk of M33. The velocity dispersion of gas was
assumed to be $c_{gas}=8$ km/s independently on the radial distance. We used the radial
distribution of gas following Corbelli (2003), allowing the contribution of helium and
heavier elements.
\pagebreak
\section{The estimates of the disk surface density}

Fig. 1 illustrates the surface density profiles of marginally stable stellar disk
calculated from the available rotation curve and the observed velocity dispersion of
planetary nebulae. The influence of gas on the stability is calculated following Rafikov,
2001) (model1) and Morozov, Khoperskov, 2005) (model 2). For comparison, the radial
density profile obtained by Ciardullo et al. (2004) is also shown.
\begin{figure}[h!]
\includegraphics[width=9cm,keepaspectratio]{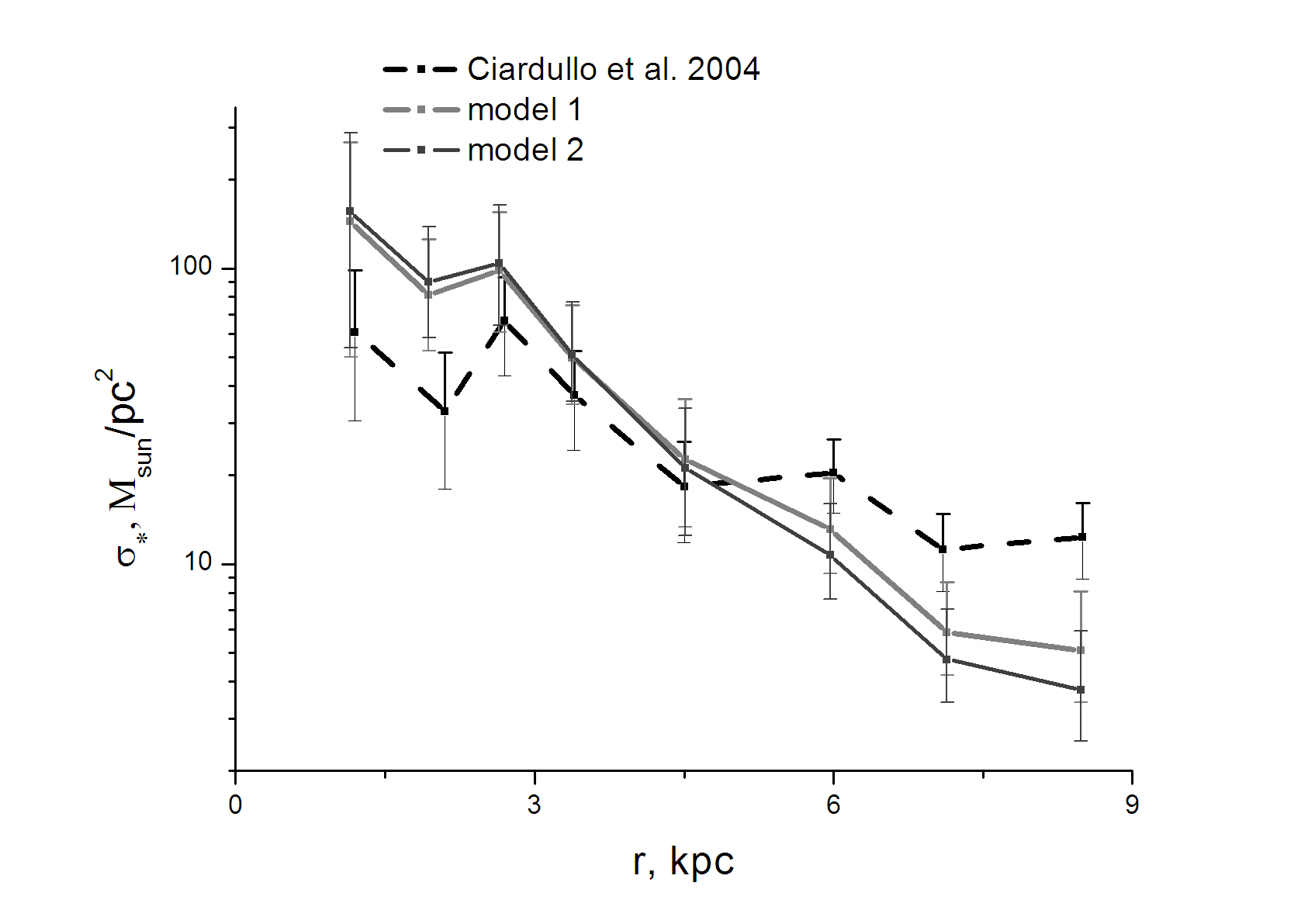}

\caption{ Radial profiles of the stellar disk surface density. Dark- and light- gray lines
demonstrate the profiles obtained for marginally stable disk taking into account the
influence of gas component on the disk stability (models 1 and 2). Dashed line denotes the
radial profile from Ciardullo et al. (2004). Here and in the following figures the error
bars correspond to the errors of velocity dispersion measurements only (Ciardullo et al.,
2004).}
\end{figure}

As it follows from Fig. 1, the difference between the models 1 and 2 is negligible, so we
restrict ourselves to the model 1 below. At the same time these profiles differ
significantly from that obtained by Ciardullo et al. (2004), where the scalelength of the
surface density distribution is about  $h\approx 4$ kpc, whereas in our models $h\approx
2$ kpc. The latter estimate is close to the photometrical radial scalelength of stellar
disk: for $K$- and $V$-bands and the accepted distance to the galaxy we have $h_K=1.45$
kpc, $h_V=2.39$ kpc (Regan,  Vogel, 1994, Baggett et al., 1998), $h_{3.6\mu m}=1.9$ kpc
(Seigar, 2011).

The inconsistence between the surface density profile of stellar disk from Ciardullo et
al. (2004) with the photometrical profile can be a result of presumption of the constant
thickness of the stellar disk.  Indeed, the thickness of a disk is not always constant
even in the first approximation. It can grow with the galactocentric distance, especially
in the disk periphery (see e.g. de Grijs, Peletier, 1997). The radial variation of stellar
disk half-thickness
$$z_{*}(r) \approx \frac{c_z^2(r)}{\pi G\sigma(r)},$$ calculated for $r < 4.5$ kpc, where
the influence of dark halo on the stellar disk thickness may be negligible, is
demonstrated in Fig. 2.

Unlike $c_r$, the  stellar disk thickness estimated from the velocity dispersion
measurements is sensitive to the choice of the badly known ratio $m=c_z / c_r$, therefore
it cannot be unambiguously determined. Fig. 2 demonstrates the radial profiles of
half-thickness $z_*(r)$ calculated for $m = 0.4$ (solid line) and  $m = 0.7$ (dash-dotted
line). As it follows from Fig. 2, the disk thickness increases rapidly to the periphery in both cases. Even for the low value $m=0.4$ it exceeds that accepted by
Ciardullo et al. (2004) beyond $r\approx$ 2.5 kpc.
\begin{figure}[h!]
\includegraphics[width=10cm,keepaspectratio]{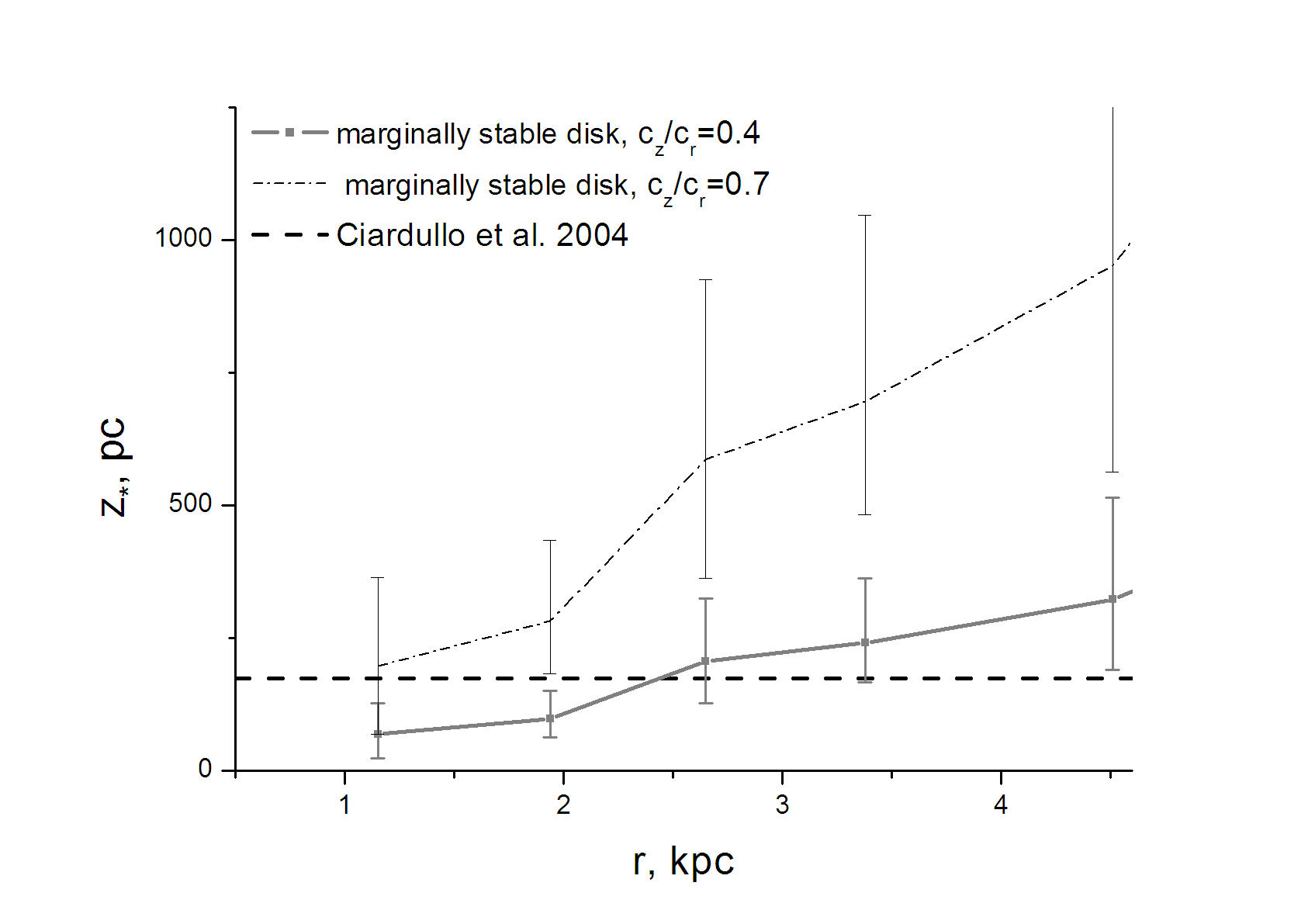}

\caption{The radial variation of the half-thickness of marginally stable stellar disk for
$c_z/c_r$=0.4 (solid line) and $c_z/c_r$=0.7 (dash-dotted line). Dashed line shows the
disk thickness adopted by Ciardullo et al. (2004).}
\end{figure}

In Fig. 3 we demonstrate the radial profiles of the local values of $M/L_K$ =
$\sigma_*/I_K$ (where $\sigma_*$ is the stellar surface density), estimated for the
marginally stable disk. The K-band surface brightness $I_K$ was taken from
Regan, Vogel (1994). Dashed line denotes $M/L_K$ profile which was obtained from the
photometric model of stellar population (Bell, de Jong, 2001)  and the observed color
indices: $(B-V)$ for the inner part (Guidoni et al. 1981) and $(H-K)$ for the outer part
of the disk (Regan, Vogel 1994). A thin dotted line in Fig. 3a corresponds to the
mass-to-light ratio corrected for internal dust extinction ($A_V\approx$ 0.25, Verley et
al. 2009). The internal extinction is negligible for $M/L_K$ ratio found from the ''red''
color index $(H-K)$ (Fig. 3b).

\begin{figure}[h!]
\includegraphics[width=8cm,keepaspectratio]{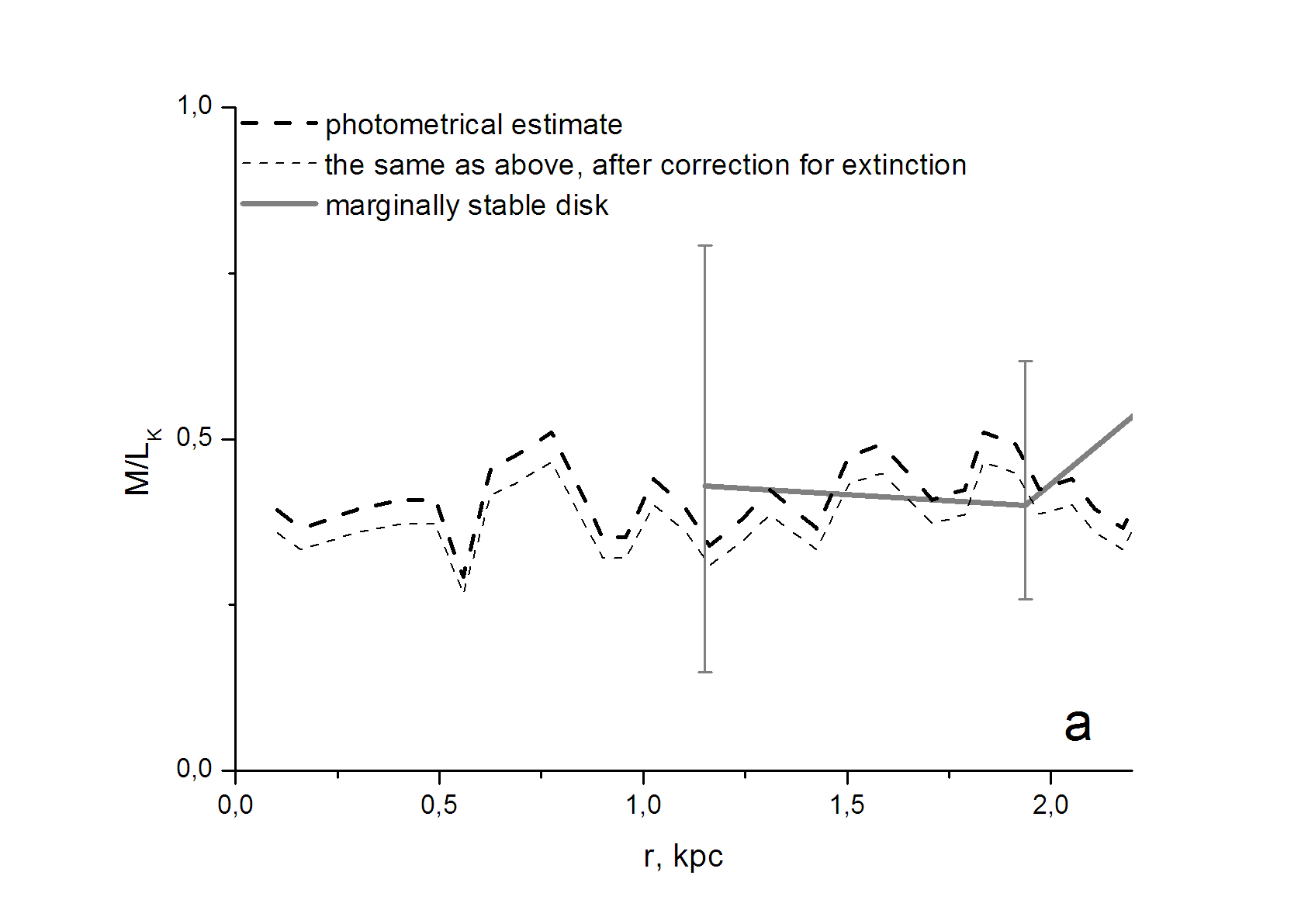}
\includegraphics[width=8cm,keepaspectratio]{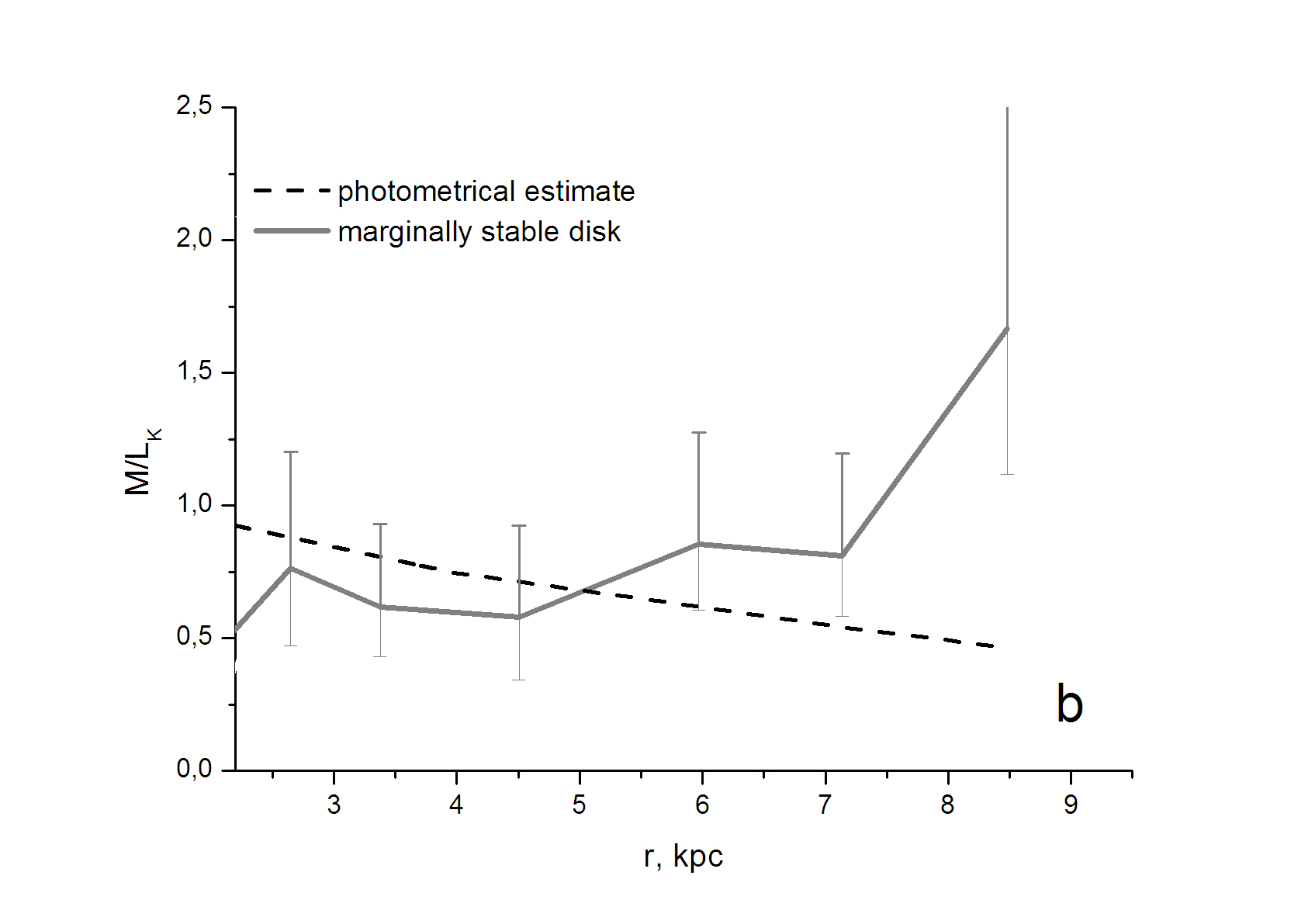}

\caption{Radial profiles of the mass-to-light ratio $M/L_K$ of marginally stable disk
(solid lines). The dashed lines correspond to the profiles based on the color indices and
stellar population synthesis models (Bell, de Jong, 2001), obtained (a) for the inner part
of the galaxy and $(B-V)$ profile from Guidoni et al. (1981); (b) for the outer part and
$(H-K)$ profile taken from Regan, Vogel 1994). Thin dotted line is the profile corrected
for the internal dust extinction $A_V=0.25$ (Verley et al., 2009). }
\end{figure}

Fig. 3 shows that the surface density of the disk, arising from the requirement of
marginal gravitational stability, is in a good agreement with the photometry-based
estimates with the exception of the most distant point ($r>7$ kpc), where the dynamical
overheating of the disk is quite possible. The real density of the disk at this point may
be twice as low as  for marginally stable condition.

The density profile of the disk obtained above was used to decompose the rotation curve,
presented by Corbelli, 2003 (Fig. 4). The model contains four components: an exponential
stellar disk with the scale length $h\approx$ 2 kpc, a gaseous disk, corresponding to H\small{I}
radial profile,  a small bulge with effective radius $r_e=2.2$ kpc (Regan, Vogel, 1994),
which has a low mass and luminosity in comparison with the disk, and a pseudo-isothermal
dark halo. Dark halo parameters were found by minimizing the deviation of the
model rotation curve from the observed one. For the dark halo mass distribution we found
 $v_{a}=167$ km/s, $a=7.5$ kpc, where $v_{a}$ is the asymptotic velocity of the halo and
$a$ is its radial scale (the core radius).
\begin{figure}[h!]
\includegraphics[width=8cm,keepaspectratio]{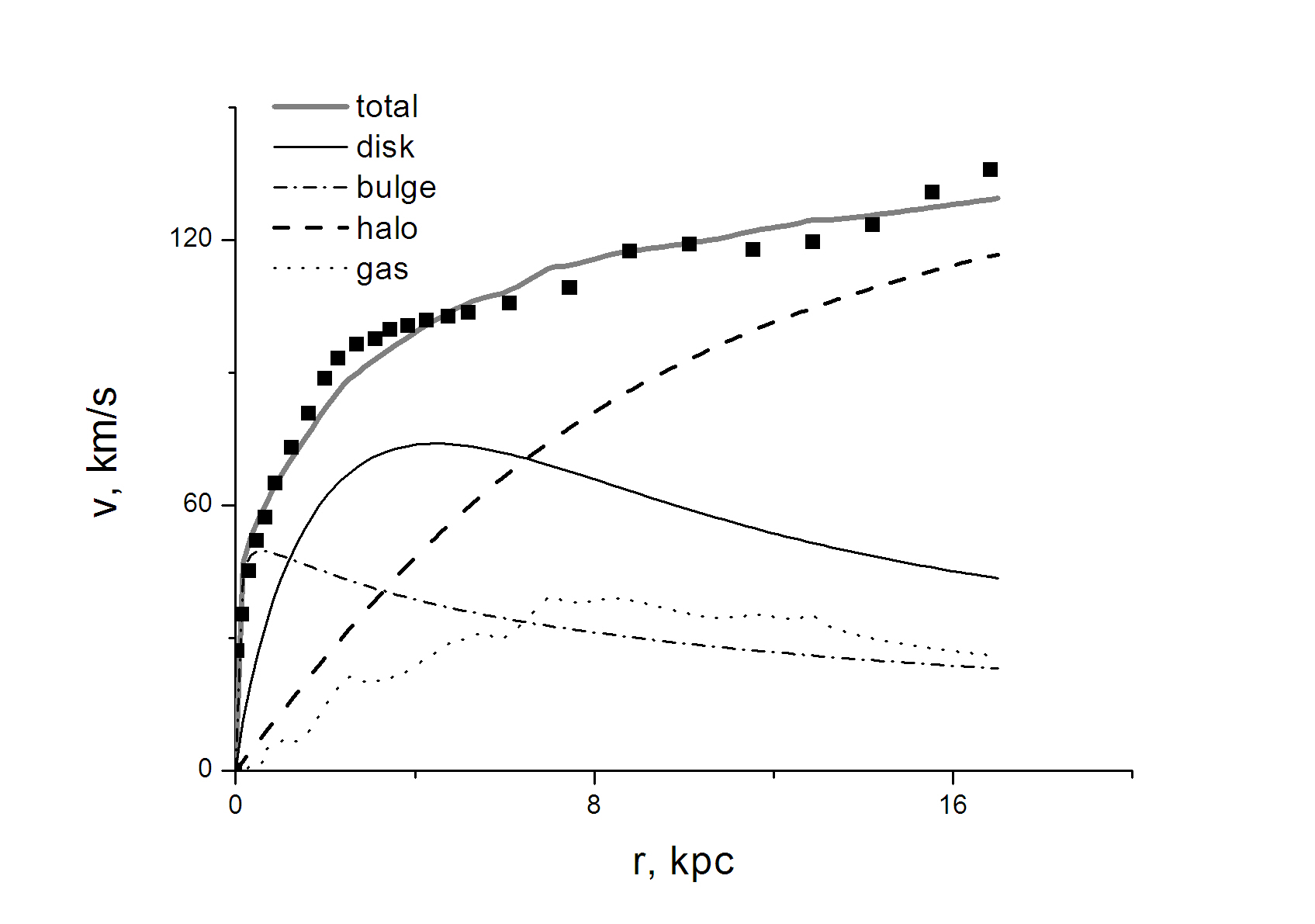}

\caption{The result of decomposition of the rotation curve taken from Corbelli, 2003
(squares) into the components: stellar and gaseous disks, bulge and dark halo, for the
marginally stable disk model.}
\end{figure}

Fig. 4 demonstrates that the marginally stable disk with the parameters we found fits well
into the observed rotation curve of the galaxy.

 Radial variation of dark halo mass to luminous (stars+gas) mass ratio
$M_h/M_{vis}$ for the model described above is shown in Fig. 5. The masses of luminous and
dark matter become equal at the galactocentric distance $r\approx 7$ kpc. Within $r=17$
kpc the dark halo mass is about 5 times higher than the mass of luminous matter -- in a
good agreement with Corbelli (2003), where the best fit model of the rotation curve was
used for a slightly lower disk scalelength $h\approx $1.45 kpc. The halo-to-disk mass
ratio inside $r=4h\approx 8$ kpc for our model ($M_h/M_{d}\approx2$) is lower than the
ratio $M_h/M_{d}\approx 3$ given in the monograph by Fridman, Khoperskov, 2011. Note that
Athanassoula et al. (1987) gave even lower dark halo-to-disk  mass ratio
$M_h/M_{d}\approx0.83$  inside of the optical radius using the spiral structure
constraints, however the uncertainty of their method is rather high (about 0.3 dex).
\begin{figure}[h!]
\includegraphics[width=9cm,keepaspectratio]{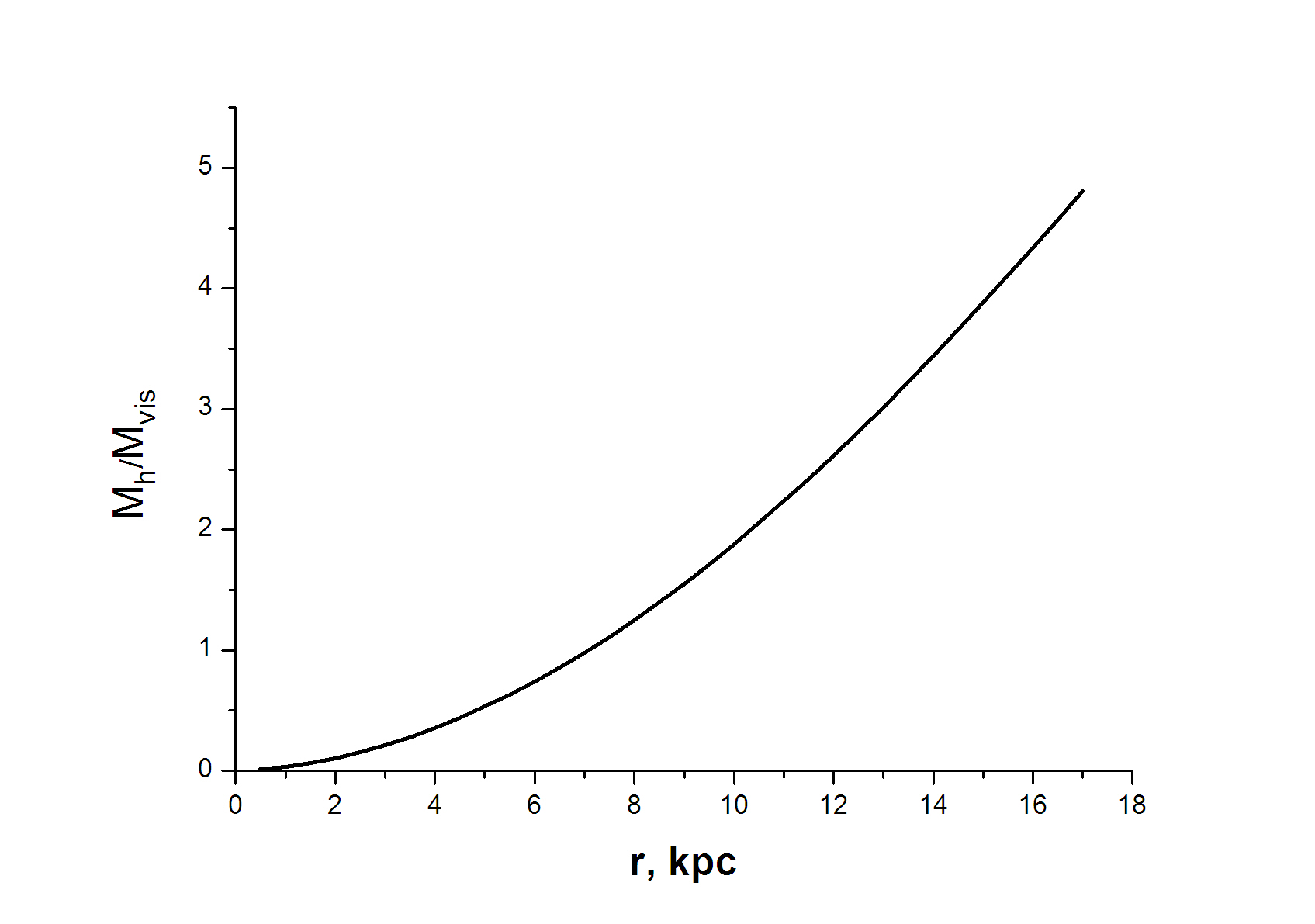}

\caption{The variation of the dark-to-luminous (stars+gas) mass ratio taken within the
radius $r$.}
\end{figure}

\section{Radial profile of oxygen effective yield in M33}
Measurements of the gas metallicity, i.e. the abundance of heavy elements produced by
stars, can give the additional information about the density of a disk and its chemical
evolution. The important parameter, which characterizes the rate of chemical enrichment,
is the yield of heavy elements $Y$, that is the mass fraction of a given chemical element
(or all heavy elements) ejected into the interstellar medium by young stars. This
parameter is connected with the nuclear evolution of stars of different masses and with
the amount of the enriched gas  that they lose. The observed abundance of any heavy
element in the interstellar gas depends both on its yield and on those processes, which,
parallel with  star formation,  affect  the mass and density of the remaining gas (such as
the blowing of gas out of the disk, the accretion of metal-poor gas, or radial gas flow).
The observations allow to estimate the so-called effective yield $Y_{eff}$, found for the
closed-box model of chemical evolution, where the following simplifications are assumed:
1) the system is closed: no gas inflow or outflow; 2) at the beginning of the disk
evolution the gas metallicity is close to zero; 3) gas is chemically homogeneous at a
given galactocentric distance; 4) stars eject the enriched gas immediately after their
formation, that is the subsequent formation of stars takes place in the already enriched
medium (the instantaneous recycling approximation); 5) the stellar initial mass function
may be considered as not-evolved. In terms of this model the abundance of heavy elements
$Z$ is defined by the yield  $Y$ and the ratio of gas mass to
 total  (gas and stars) mass of a disk:
\begin{equation}
Z = Y \cdot{ln(1/\mu)},
\end{equation}
 where $\mu=\sigma_{gas}(r)/\sigma$(r) is the gas mass fraction of a disk at a given
 galactocentric distance. The accretion of metal-poor gas as well as the outflow of gas
 leads to the decrease of $Y_{eff}$ making it lower than the
 real yield $Y$ (Edmunds, 1990). Hence, the simple model of chemical evolution
 of a galaxy is useful for testing the assumptions  which it is based on.

The substantial fraction of the enriched matter ejected by stars belongs to oxygen, the
most abundant element after $H$ and $He$. The main producers of oxygen are the most
massive short-lived stars, for which the condition of instantaneous recycling satisfies
best of all. Thus it is very convenient to refer $Y_{eff}$ to the effective yield of this
particular element:
\begin{equation}
Y_{eff}=\frac{12(O/H)}{ln(1/\mu)},
\end{equation}
where $12(O/H)$ is the mass fraction of oxygen. The known radial distributions of the disk
surface density of M33 and $O/H$ ratio allow to calculate the radial profile $Y_{eff}(R)$
and compare it to those predicted by different scenarios of the disk evolution. To
estimate $\mu$ we used the radial distribution of hydrogen density taken from Corbelli
(2003), applying the coefficient 1.33 for the total gas density.

Oxygen abundance $O/H$ at different galactocentric distances of M33  was estimated by
different authors. A significant spread of $O/H$ values for HII regions, even for those
that are at the same distance from the center, as well as some anomalies of oxygen
abundance of giant HII regions in the central part of M33, led to a significant divergence
of the existing estimates of radial gradient of $O/H$ (see the discussion in Magrini et
al., 2010). Furthermore, until recently there were only a small number of HII regions with
the reliable measurements of $O/H$ obtained from the gas temperature estimates $T_e$.
Pilyugin et al. (2004) applied the developed P-method, which agrees with the $T_e$-based
method, to the available spectra of HII regions in the sample of galaxies. For M33 they
obtained the radial gradient of $O/H$ $\approx$ -0.20 dex/$r_{25}$, which corresponds to
-0.02 dex/kpc. The most complete list of $O/H$ data for HII regions with known $T_e$ was
given later by Magrini et al. (2010). The gradient of $O/H$ they found is -0.033 $\pm$
0.008 dex/kpc (for the distance accepted here). Below we use both gradients to calculate
the radial distribution of $Y_{eff}$.

 The results are demonstrated in Fig. 6a,b. Different lines correspond to the different
 radial density distribution. Thick solid line is related to the model of marginally
 stable disk. Dash-dotted and dashed
 lines correspond to the photometrically estimated density (see
 above) and to the model of Ciardullo et al. (2004). Thin horizontal line $Y_{eff}=0.035$
 denotes the estimate of real oxygen yield $Y_O$ found by Pilyugin et al. (2007) from
 the gas metallicity in the central regions of spiral galaxies
 with the highest oxygen abundances.  Close estimates of oxygen yield were obtained earlier by Bresolin et al.
 (2004) ($Y_{O}$=0.032) and Pilyugin et al. (2004) ($Y_{O}$=0.027).

\begin{figure}[h!]
\includegraphics[width=8cm,keepaspectratio]{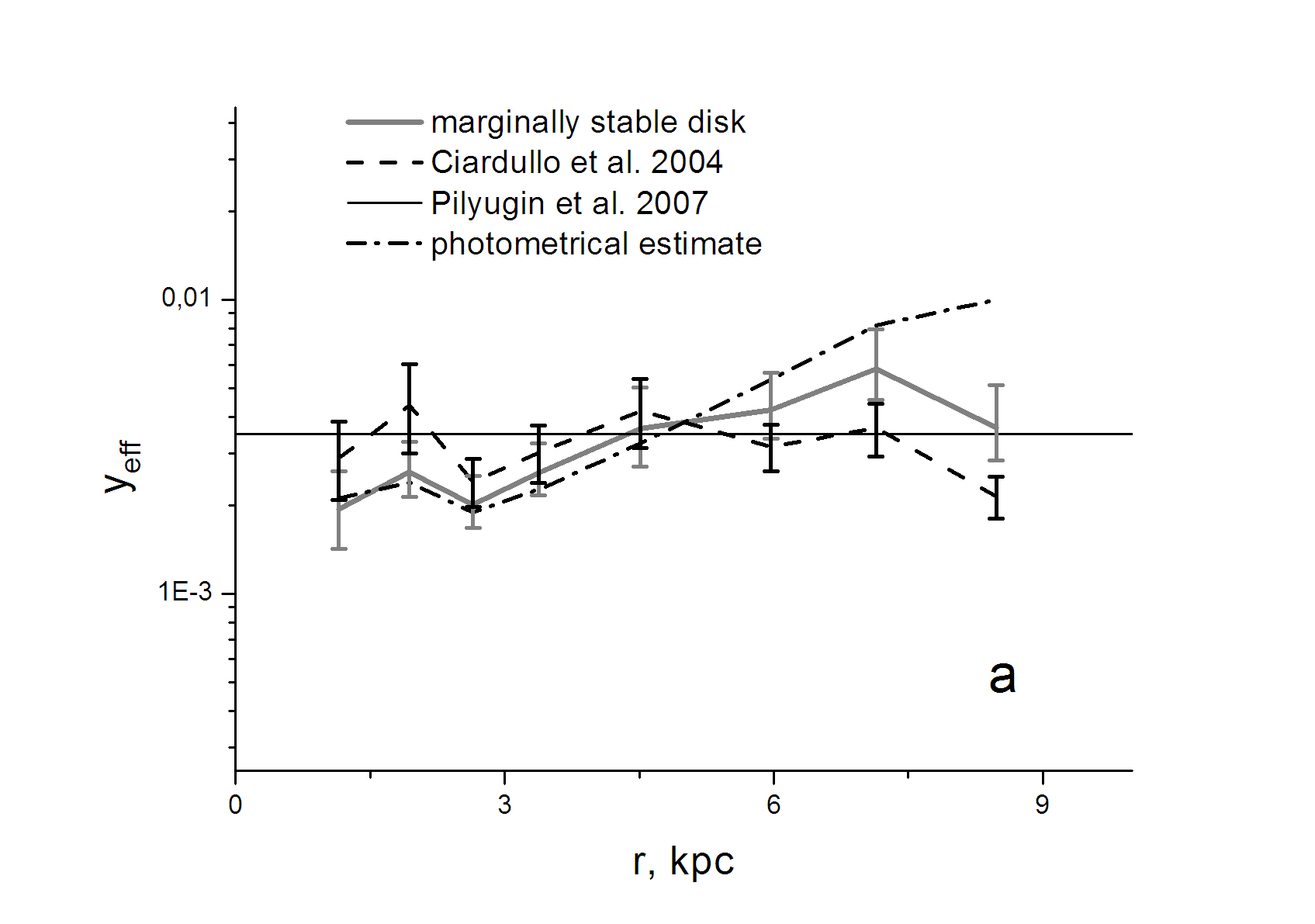}
\includegraphics[width=8cm,keepaspectratio]{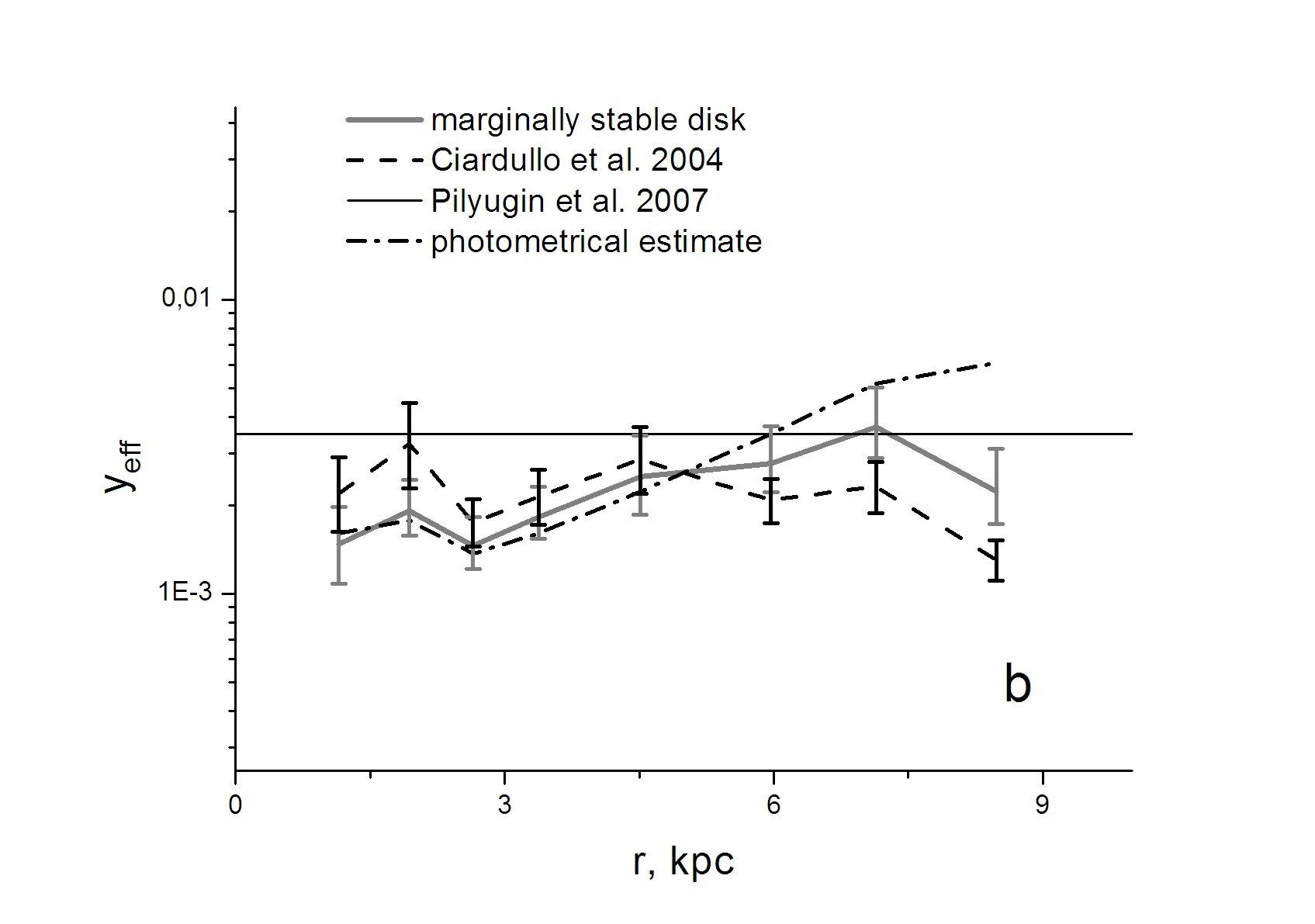}

\caption{Radial distribution of the effective yield calculated from the radial profile of
$O/H$ from Pilyugin et al., 2004 (a) and Magrini et al., 2010 (b). Solid line is for the
marginally stable disk model, dash-dotted line corresponds to the photometrically obtained
surface density of the disk,  dashed line is for the disk surface density taken from
Ciardullo et al. (2004). The horizontal line marks the oxygen yield $Y_O$ found by
Pilyugin et al. 2007 for the gas in the central regions of spiral galaxies with the
highest oxygen abundance.}
\end{figure}

It follows from Fig. 6, that the effective yield $Y_{eff}$ increases from the center to
the disk periphery for all radial profiles of the surface density we considered except the
profile of Ciardullo et al. (2004), where $Y_{eff}$ is approximately constant. Note
however, that $Y_{eff}=const$ would mean that the gas accretion onto the galaxy is absent
or at least it does not affect the present day abundance.  It badly agrees with the
current models of chemical evolution which include the external accretion as the necessary
ingredient to account for the observed distribution of metallicity and density of gas in
M33 (Magrini et al. 2007, 2010). Low $Y_{eff}$ for the last point ($r\approx $ 8 kpc), if
it is real, may be caused by the density overestimation due to the dynamical overheating
of the disk at large radii, as it was noted above, because the photometry-based model does
not reveal this feature. A reduction of $Y_{eff}$ toward the center may indicate that the
 accretion of metal-poor gas is more essential for the inner part of the galaxy. In principle,
 it may be attributed either to the external accretion described by the current models of chemical
evolution, or to the internal accretion, i.e. to the radial drift of less enriched gas
toward the center. It is remarkable that the $O/H$ data taken from Magrini et al. (2010),
which are probably the most reliable, give the estimate of $Y_{eff}$ at large radial
distance close to $Y_O$, expected in the absence of accretion (the horizontal line in Fig.
6).

\section{Conclusions}
In summary, the assumption that the stellar velocity dispersion in the disk of M33 inside
of $r\sim 6-7$ kpc is close to the minimal value needed to stabilize it, does not
contradict either the photometrical profile of the disk, or the  $M/L$ ratio of stellar
population, or the rotation curve of the galaxy, or the radial distribution of oxygen in
the disk. It means that the disk of this galaxy within several radial scalelengths have
not experienced a significant dynamical heating caused by interaction with nearby galaxies
of Local Group or by minor merging events  during its evolution. The existence of the
massive dark halo is also confirmed: the dark halo mass begins to dominate over the disk
mass starting with the radius of about 7 kpc.  The radial distribution of the effective
yield of oxygen, calculated in the frame of marginally stable disk model, decreases to
the center. It supports the conclusion followed from the current chemical evolution models
about the significant role of accretion in the chemical enrichment of interstellar medium
in the inner disk. However these conclusions may not be extrapolated to the outer regions
of the disk at $r>7~ -~8$ kpc where the dynamical overheating of the disk is
quite possible and the dynamical and chemical evolution history may be  more complicated.\\
\bigskip
\bigskip

This work was supported by Russian Foundation for Basic Research, grant 11-02-12247.
\pagebreak

\end{document}